\documentclass[lettersize,journal]{IEEEtran}
\usepackage{amsmath,amsfonts,amssymb}
\usepackage{algorithmic}
\usepackage{algorithm}
\usepackage{array}
\usepackage[caption=false,font=normalsize,labelfont=sf,textfont=sf]{subfig}
\usepackage{textcomp}
\usepackage{stfloats}
\usepackage{url}
\usepackage{verbatim}
\usepackage{graphicx}
\usepackage{cite}
\usepackage{xcolor}
\definecolor{forestgreen}{rgb}{0.133, 0.545, 0.133}
\usepackage{pgf}
\usepackage{pgfplots}
\usepackage{tikz}
\usepackage{eqparbox}
\usepackage{makecell}
\newcommand{\norm}[1]{\left\Vert#1\right\Vert}

\newcommand{\complexm}[2]{\mathbb{C}^{#1\times #2}}

\newcommand{\CN}[2]{\mathcal{CN}{(#1,#2)}}

\newcommand{\trace}[1]{{\rm Tr}\left(#1\right)}
\newcommand{\real}{\mathcal{R}}

\begin{document}

\title{Decentralized Algorithms for Out-of-System Interference Suppression in Distributed MIMO}
% Decentralized algorithms for suppression and managing of out-of-system interference for distributed MIMO  in uplink-downlink

\author{Zakir Hussain Shaik, \IEEEmembership{Graduate Student Member, IEEE}, and Erik G. Larsson \IEEEmembership{Fellow, IEEE}
\thanks{This work was partially funded by the REINDEER project of the European Union‘s Horizon 2020 research and innovation program under grant agreement No. 101013425. This work was also partially supported by ELLIIT and KAW.}}



\maketitle
\begin{abstract}
Out-of-system (OoS) interference is a potential limitation for distributed networks that operate in unlicensed spectrum or in a spectrum sharing scenario. The OoS interference differs from the in-system interference in that OoS signals and their associated channels (or even their statistics) are completely unknown. In this paper, we propose a novel distributed algorithm that can mitigate OoS interference in the uplink and suppress the signal transmission in the OoS direction in the downlink. To estimate the OoS interference, each access point (AP), upon receiving an estimate of OoS interference from a previous AP, computes a better estimate of OoS interference by rotate-and-average using Procrustes method and forwards the estimates to the next AP. This process continues until the central processing unit (CPU) receives the final estimate. Our method has comparable performance to that of a fully centralized  interference rejection combining algorithm and has much lower fronthaul load requirements.
\end{abstract}

\vspace{-4mm}

\section{Introduction}
%Cell-free massive multiple-input-multiple-output (MIMO), with its high spectral efficiency and macro-diversity features forms a suitable candidate for many future applications, especially in unlicensed spectrum or spectrum sharing scenarios\cite{parkvall20205g}. While this fosters opportunities for new applications, it also introduces new signal processing challenges \cite{zhang2017survey,ahmad20205g}. One challenge is the presence of out-of-system (OoS) interfering sources. Unlike in-system interference caused by other UEs, the OoS interference differs in that its signal components are completely unknown, i.e., they may or may not contain pilots, and even if they are present, these are unknown. Therefore, efficient methods to reject OoS interference are crucial in such applications.%, especially in  cell-free massive MIMO where the interference affects multiple distributed nodes through different channels.

As wireless standards are evolving, there is an increase in demand for applications in unlicensed spectrum and also employing distributed networks such as cell-free massive multiple-input-multiple-output (MIMO). In such applications, one major challenge is the presence of the presence of out-of-system (OoS) interfering sources and need for efficient methods to reject the OoS interference. In the literature, the most commonly employed interference rejection method is to perform decoloring of the received signal to prewhitening (by obtaining a sample covariance of the interference plus noise). Then the desired signals of the serving UEs are estimated from the prewhitened received signal \cite{karlsson1996interference,winters1993signal,kuzminskiy2021maximum,yang2022jamming,davydov2020robust}. This approach is commonly referred as interference rejection combining (IRC). Most of the works either process all the signals centrally, or in a decentralized manner at each AP locally without cooperation among the APs. The former approach suffers from significant fronthaul load; for instance with a stripe topology, to accumulate all signals at the central node. The latter approach does not exploit the fact that the OoS interfering signal is the same across all APs.

{\it Motivation and Main Contributions of the Paper:} For the applications in unlicensed spectrum, e.g., WiFi,  OoS 
 may be a significant interference source. Effectively suppressing and managing  OoS interference sources at the same time as serving the users will be crucial in such applications. Some  challenges when dealing with OoS interference are: no prior information about the OoS source is available; not even its signal statistics; the interference affects different APs differently through different channels
 (yet, cancelling the OoS interference requires phase-coherent processing among the APs); and the interference-suppressing algorithms should ensure that the fronthaul load on the links between the APs remains constant irrespective of the number of  APs. The cell-free massive MIMO with in-system interference is a well investigated problem \cite{cellfreebook}. However, the presence of OoS interfering sources have not been adequately investigated for cell-free massive MIMO. To the best of our knowledge, existing literature did not address this problem except in the case of single OoS source with daisy-chain topology \cite{shaik2023distributed}. In this paper, we first provide a novel distributed algorithm to estimate the OoS interference. The specific contributions of the paper are: we propose a novel distributed processing algorithm using the orthogonal Procrustes problem to estimate and suppress multiple OoS interference sources, we provide a Gramian based algorithm to handle multiple OoS sources with superior performance but also relatively large fronthaul load. We demonstrate the effectiveness of the OoS interference estimates in the uplink and downlink payload phase. In the uplink, we treat OoS sources as additional fictitious UEs and similarly in the downlink we nullify the signals in the direction of the OoS sources. Moreover, the proposed methods do not assume any specific properties of the channels; in particular, the methods do not rely on  channel hardening or favorable propagation.\vspace{-2mm}

\section{System Model and Problem Formulation}
We consider a cell-free massive MIMO network comprising $L$ APs each equipped with $N$ antennas serving $K$ UEs in the presence of $K_{\mathcal{I}}$ single-antenna OoS interference sources.\footnote{Alternatively, a single OoS source with $K_{\mathcal{I}}$ antennas. Thus, our model \eqref{recSigPilotPhase} naturally extends to multiple OoS sources with multiple antennas (say $M$) by replacing $\mathbf{G}_l\mathbf{S}^H$ to $\sum_{j=1}^{K_{\mathcal{I}}}\mathbf{G}_{lj}\mathbf{S}_j^H$, where $\mathbf{S}_j\in \complexm{\tau_p}{M}$ and $\mathbf{G}_{lj}$ are OoS $j$ transmit signal and its channel to the AP $l$, respectively.} We consider a block fading channel between the UEs (plus OoS sources) and the APs. Further, we assume the system operates in time-division duplex (TDD) mode, leveraging channel reciprocity. This facilitates downlink beamforming design based on uplink channel estimates.

During the channel estimation phase, we assume that there are $K$ orthogonal pilots of length $\tau_p \geq K + K_{\mathcal{I}}$. We denote the pilot vector of UE $k$ by $\boldsymbol{\phi}_k$, normalized such that $\norm{\boldsymbol{\phi}_k} = 1$. The signal received at AP $l \in [L] \triangleq \left\{1,\ldots,L\right\}$ during the pilot phase of $\tau_p$ channel uses is
\begin{equation}\label{recSigPilotPhase}
	\mathbf{Y}_l = \sqrt{\rho\tau_p}\mathbf{H}_l\boldsymbol{\Phi}^H + \mathbf{G}_l\mathbf{S}^H + \mathbf{N}_l,
\end{equation}
where $\rho$ is the transmit signal-to-noise ratio (SNR) of UEs, $\mathbf{H}_l = [\mathbf{h}_{1l},\ldots,\mathbf{h}_{Kl}]$, where $\mathbf{h}_{kl} \in \complexm{N}{1}$ is the channel between UE $k$ and AP $l$, $\mathbf{G}_l = [\mathbf{g}_{1l},\ldots,\mathbf{g}_{K_{\mathcal{I}}l}]$ where $\mathbf{g}_{jl} \in \complexm{N}{1}$ is the channel between OoS source $j$ to AP~$l$, $\mathbf{S} = [\mathbf{s}_{1},\ldots,\mathbf{s}_{K_{\mathcal{I}}}]$ where $\mathbf{s}_{j}\in \complexm{\tau_p}{1}$ is the signal from OoS source $j\in [J]$, pilot matrix $\boldsymbol{\Phi} = [\boldsymbol{\phi}_1,\ldots,\boldsymbol{\phi}_K]$ and $\mathbf{N}_l\in \complexm{N}{\tau_p}$ is the noise matrix at AP $l$ where all entries are i.i.d. $\CN{0}{1}$. \vspace{-2mm}

\subsection{Uplink Payload Model}
In the uplink payload phase, in each symbol period AP $l$ receives the signal
\begin{equation}\label{ulSigMod}
	\begin{aligned}
		\mathbf{y}_l^{\rm ul} &= \mathbf{H}_l\mathbf{x}^{\rm ul} + \mathbf{G}_l
		\mathbf{s} + \mathbf{n}_l^{\rm ul}=\begin{bmatrix}
			{\mathbf{H}}_l & {\mathbf{G}}_l
		\end{bmatrix} \begin{bmatrix}
			\mathbf{x}^{\rm ul}\\ \mathbf{s}
		\end{bmatrix} + \mathbf{n}_l^{\rm ul},\\
	\end{aligned}
\end{equation}
where $\mathbf{x}^{\rm ul}\in \complexm{K}{1}$ is the signal collectively transmitted by the $K$ UEs, $\mathbf{s}\in \complexm{K_{\mathcal{I}}}{1}$ is the signal from the $K_{\mathcal{I}}$ OoS sources and $\mathbf{n}_l^{\rm ul}~\sim~\CN{\mathbf{0}}{\mathbf{I}_N}$ is the noise vector at AP $l$, where $\mathbf{I}_N$ is the identity matrix of dimension $N\times N$. \vspace{-3mm}

\subsection{Downlink Payload Model}
The downlink signal collectively received by all the UEs in a symbol period is given by
\begin{equation}\label{dlSigMod}
	\mathbf{y}^{\rm dl} = \sum_{l=1}^{L}\mathbf{H}_l^H\mathbf{W}_l\mathbf{x}^{\rm dl} + \mathbf{n}^{\rm dl},
\end{equation}
where $\mathbf{x}^{\rm dl}\in \complexm{K}{1}$ contains the signals destined to the UEs. Specifically, the $k$th entry of $\mathbf{x}^{\rm dl}$ and $\mathbf{y}^{\rm dl}$ are the transmit and received signal of UE $k$. Further, $\mathbf{W}_l$ is the precoding matrix at AP $l$. We assume $\mathbf{n}^{\rm dl}\sim \CN{\mathbf{0}}{\mathbf{I}_K}$. \vspace{-2mm}

\subsection{Motivation and Problem Statement}
Our primary goal is to suppress the OoS interference while providing services in the uplink and the downlink. To do that, we want to estimate $\{\mathbf{G}_l\}, l\in [L]$ which then can be used for uplink receive combining phase and downlink precoding. To make a good estimate of $\{\mathbf{G}_l\}, l\in [L]$, we need a good estimate of the interfering signal $\mathbf{S}$, which is the same across all the APs. In the subsequent sections, we propose a methodology to estimate OoS interference which is applicable to any distributed topology such as a tree, mesh or daisy-chain. However, to keep the exposition of the proposed algorithms simpler, we will consider a daisy-chain topology as the underlying topology for the discussion to follow in the rest of the paper. \vspace{-2mm}

\section{Channel and OoS Interference Estimation}\label{sec:chanOoSesti}
Channel estimation is done locally at each AP. We assume that each AP employs least-squares (LS) estimation and accordingly the estimate of the channel at AP $l$ in a coherence interval is given by
\begin{equation}\label{channEst}
	\begin{aligned}
		\widehat{\mathbf{H}}_l &= \frac{1}{\sqrt{p\tau_p}}\mathbf{Y}_l\boldsymbol{\Phi} = \mathbf{H}_l + \frac{1}{\sqrt{p\tau_p}}\left(\mathbf{G}_l\mathbf{S}^H + \mathbf{N}_l\right)\boldsymbol{\Phi}.
	\end{aligned}
\end{equation}

Having estimated the UEs channels using \eqref{channEst} at all APs, we now focus on estimating the channels and signals of the OoS sources. As a first step, each AP obtains the following residual signal that effectively captures the OoS interference plus noise at AP $l$:
\begin{equation}
	\begin{aligned}
		\mathbf{Z}_l &=  \mathbf{Y}_l - \sqrt{p\tau_p}\widehat{\mathbf{H}}_l\boldsymbol{\Phi}^H = \left(\mathbf{G}_l\mathbf{S}^H + \mathbf{N}_l\right)\boldsymbol{\Pi}_{\boldsymbol{\Phi}}^{\perp},
	\end{aligned}
\end{equation}
where $\boldsymbol{\Pi}_{\boldsymbol{\Phi}}^{\perp} = \mathbf{I}_{\tau_p} - \boldsymbol{\Phi}\boldsymbol{\Phi}^H$ is the projection matrix onto the orthogonal complement of $\boldsymbol{\Phi}$. To obtain a reasonable estimate of the interfering signal $\mathbf{S}$ from the residual matrices, we require $\tau_p \geq K + K_{\mathcal{I}}$, otherwise no degrees of freedom are left in $\boldsymbol{\Pi}_{\boldsymbol{\Phi}}^{\perp}$ i.e., the projection matrix would be a zero matrix. Further, note that we can estimate $\mathbf{S}$ up to $\tau_p-K$ dimensions and this is because $\boldsymbol{\Pi}_{\boldsymbol{\Phi}}^{\perp}$ is a projection matrix with rank $\tau_p - K$. Therefore, we instead focus on estimating the projected component of $\mathbf{S}$. For that we first decompose the projection matrix using the economy-size singular value decomposition as
\begin{equation}
	\boldsymbol{\Pi}_{\boldsymbol{\Phi}}^{\perp} = \boldsymbol{\Psi}\boldsymbol{\Psi}^H,
\end{equation}
where $\boldsymbol{\Psi} \in \complexm{\tau_p}{(\tau_p - K)}$ satisfies $\boldsymbol{\Psi}^H\boldsymbol{\Psi}  = \mathbf{I}_{\tau_p-K}$. We denote the signal we aim to estimate by $\bar{\mathbf{S}} = \boldsymbol{\Psi}^H\mathbf{S}$,
%\begin{equation}
%	
%\end{equation}
which contains the coordinates of $\mathbf{S}$ in the basis given by $\mathbf{\boldsymbol{\Psi}}$.

To obtain an estimate $\widehat{\bar{\mathbf{S}}}$ of $\bar{\mathbf{S}}$, we work on a lower dimensional residual signal instead, which is as follows:
\begin{equation}
	\begin{aligned}
		\mathbf{Z}_l\boldsymbol{\Psi} &= \left(\mathbf{G}_l\mathbf{S}^H + \mathbf{N}_l\right)\boldsymbol{\Psi} = \mathbf{G}_l\bar{\mathbf{S}}^H + \mathbf{N}_l^{'},
	\end{aligned}
\end{equation} 
where $\mathbf{N}_l^{'} = \mathbf{N}_l\boldsymbol{\Psi}\in \complexm{N}{(\tau_p - K)}$ is a noise matrix whose entries are i.i.d. because $\boldsymbol{\Psi}$ is a unitary matrix. In the following subsection, we present two methods, with different fronthaul-performance trade-off, to estimate the OoS signals $\bar{\mathbf{S}}$ and their corresponding channels $\{\mathbf{G}_l\}$ using locally obtained processed residual signals $\left\{\mathbf{Z}_l\boldsymbol{\Psi}\right\}$.

In both methods, we give a decentralized solution to the following centralized LS problem of estimating $\bar{\mathbf{S}}$:
\begin{equation} \label{optForm1}
	\underset{\mathbf{G},\bar{\mathbf{S}}}{\rm{minimize}} \quad  \norm{\mathbf{Z}\mathbf{\Psi} - \mathbf{G}\bar{\mathbf{S}}^H}_F,
\end{equation}
where $\mathbf{Z} = [\mathbf{Z}_1^T,\ldots,\mathbf{Z}_L^T]^T$, $\mathbf{G}= [\mathbf{G}_1^T,\ldots,\mathbf{G}_L^T]^T$ and $\norm{\cdot}_F$ is the Frobenius norm. The cost of solving \eqref{optForm1}  using the SVD is approximately $\mathcal{O}(K_{\mathcal{I}}NL\tau_p )$ floating-point operations (flops)~\cite{halko2011finding}.\vspace{-4mm}

\subsection{Method 1: Sequential Unitary Rotation and Averaging}
In this method, we exploit the fact that $\bar{\mathbf{S}}$ is the same at all APs. In the first step, each AP makes a {\em local} estimate of $\bar{\mathbf{S}}$, which we denote by $\widehat{\bar{\mathbf{S}}}_l^{\circ},\ l \in [L]$. This local estimate is obtained by solving the following local LS problem
\begin{equation} \label{optForm1_Local}
	\underset{\mathbf{G}_l,\bar{\mathbf{S}}}{\rm{minimize}} \quad  \norm{\mathbf{Z}_l\mathbf{\Psi} - \mathbf{G}_l\bar{\mathbf{S}}^H}_F.
\end{equation}
One way to obtain a global solution of \eqref{optForm1_Local} is to take the best rank-$K_{\mathcal{I}}$ approximation of $\mathbf{Z}_l\mathbf{\Psi}$ using SVD, where the estimate of $\bar{\mathbf{S}}$, $\widehat{\bar{\mathbf{S}}}_l^{\circ}$, is the right singular matrix, and the estimate of $\mathbf{G}_l$ is the left singular matrix scaled by the diagonal matrix containing singular values. 

Although $\bar{\mathbf{S}}$ is the same across all APs, the SVD estimates of it at all APs need not be the same and can be ambiguous up to an arbitrary unitary rotation. Therefore, simply averaging the estimates APs is not a wise approach. To address this problem, we propose that AP $l$ upon receiving the estimate, $\widehat{\bar{\mathbf{S}}}_{l-1}$, from AP $l-1$ should compute its 
estimate as follows:\footnote{The connection order of the APs impacts performance, making the choice of the initiating AP, for a given  interconnection topology, an intriguing open problem.}
\begin{equation}\label{RotnAvg}
	\widehat{\bar{\mathbf{S}}}_l = 0.5\left(\widehat{\bar{\mathbf{S}}}_{l-1} + \widehat{\bar{\mathbf{S}}}_l^{\circ}\mathbf{Q}_l^H\right),
\end{equation}
where $\mathbf{Q}_l \in \complexm{K_{\mathcal{I}}}{K_{\mathcal{I}}}$ is a unitary matrix that rotates $\widehat{\bar{\mathbf{S}}}_l^{\circ}$ to minimize its distance to $\widehat{\bar{\mathbf{S}}}_{l-1} $. Then, the AP $l$ forwards the estimate, \eqref{RotnAvg} to AP $l+1$. In \eqref{RotnAvg}, the matrix $\mathbf{Q}_l$ is the rotation matrix that aligns the estimates $\widehat{\bar{\mathbf{S}}}_{l-1}$ and $\widehat{\bar{\mathbf{S}}}_l^{\circ}$ in the LS sense. Mathematically we obtain $\{\mathbf{Q}_l\}$ as a solution to the following orthogonal {\it Procrustes} problem \cite{schonemann1966generalized}:
\begin{equation} \label{procrustesProb1}
	\underset{\mathbf{Q}_l}{\rm{minimize}} \quad  \norm{\widehat{\bar{\mathbf{S}}}_l^{\circ}\mathbf{Q}_l^H - \widehat{\bar{\mathbf{S}}}_{l-1}}_F.
\end{equation}

Fortunately, we can obtain $\mathbf{Q}_l$ in semi-closed form. To do that, first we obtain the SVD of $\widehat{\bar{\mathbf{S}}}_l^{\circ H}\widehat{\bar{\mathbf{S}}}_{l-1} = \mathbf{U}_l \boldsymbol{\Lambda}_l \mathbf{V}_l^H$, where we denote $ \mathbf{V}_l$, $ \mathbf{U}_l$ and $\boldsymbol{\Lambda}_l$ to be its right singular matrix, left singular matrix and diagonal matrix with singular values, respectively. Then, we obtain the optimal solution to \eqref{procrustesProb1} as follows:
\vspace{-3mm}
\begin{equation} \label{procrustesProof}
	\begin{aligned}
		\underset{\mathbf{Q}_l:\mathbf{Q}_l^H\mathbf{Q}_l = \mathbf{I}}{\rm{minimize}} &\quad  \norm{\widehat{\bar{\mathbf{S}}}_l^{\circ}\mathbf{Q}_l^H - \widehat{\bar{\mathbf{S}}}_{l-1}}_F\\
	\implies	\underset{\mathbf{Q}_l:\mathbf{Q}_l^H\mathbf{Q}_l = \mathbf{I}}{\rm{maximize}} &\quad  \real\left\{\trace{\mathbf{Q}_l \widehat{\bar{\mathbf{S}}}_l^{\circ H}\widehat{\bar{\mathbf{S}}}_{l-1}}\right\},\\
		%\implies	\underset{\mathbf{Q}_l:\mathbf{Q}_l^H\mathbf{Q}_l = \mathbf{I}}{\rm{maximize}} &\quad  \real\left\{\trace{\mathbf{V}_l^H\mathbf{Q}_l \mathbf{U}_l\boldsymbol{\Lambda}_l}\right\},\\
		%\implies	\underset{\widetilde{\mathbf{Q}}_l: \widetilde{\mathbf{Q}}_l^H\widetilde{\mathbf{Q}}_l = \mathbf{I}}{\rm{maximize}} &\quad  \real\left\{\trace{\widetilde{\mathbf{Q}}_l\boldsymbol{\Lambda}_l}\right\}, \\
		\implies	\underset{\left\{\widetilde{\mathbf{Q}}_l[i,i]\right\}}{\rm{maximize}} &\quad  \sum_{i=1}^{K_{\mathcal{I}}}\real\left\{\widetilde{\mathbf{Q}}_l[i,i]\right\}\boldsymbol{\Lambda}_l[i,i],\\
		\overset{(a)}{\implies} \widetilde{\mathbf{Q}}_l &= \mathbf{I}_{K_{\mathcal{I}}},\\
		\implies \mathbf{Q}_l &= \mathbf{V}_l\mathbf{U}_l^H,
		\end{aligned}
\end{equation}
where $\widetilde{\mathbf{Q}}_l \triangleq \mathbf{V}_l^H\mathbf{Q}_l \mathbf{U}_l$, $(a)$ follows from observation that $\widetilde{\mathbf{Q}}_l$ is unitary, implying that its elements are bounded by unity and thus the maximum is achieved when diagonal entries are equal to one. For the special case $K_{\mathcal{I}} = 1$, $\mathbf{Q}_l=e^{i\theta_l} (\text{scalar}), \theta_l \in [0,2\pi]$ and the problem reduces to that discussed in \cite{shaik2023distributed}:
\begin{equation}
\underset{\theta_l}{\rm{minimize}} \quad  \norm{\widehat{\bar{\mathbf{s}}}_l^{\circ}e^{-i\theta_l} - \widehat{\bar{\mathbf{s}}}_{l-1}}_2.
\end{equation}

After the CPU obtains the final estimate $\widehat{\bar{\mathbf{S}}} = \widehat{\bar{\mathbf{S}}}_{L}$, it forwards the estimate to all the APs. Then, each AP makes a local channel estimate by solving
\begin{equation}
	\underset{\mathbf{G}_l}{\rm{minimize}} \quad  \norm{\mathbf{Z}_l\mathbf{\Psi} - \mathbf{G}_l\widehat{\bar{\mathbf{S}}}^H}_F,
\end{equation}
which is given by
\begin{equation}\label{eqn:intChanEst}
	\widehat{\mathbf{G}}_l = \mathbf{Z}_l\mathbf{\Psi}\widehat{\bar{\mathbf{S}}}(\widehat{\bar{\mathbf{S}}}^H\widehat{\bar{\mathbf{S}}})^{-1}.
\end{equation}

The fronthaul load in each link between the APs is $2K_{\mathcal{I}}(\tau_p - K)$ real symbols per link. The cost of solving \eqref{RotnAvg} is approximately $\mathcal{O}(\tau_pNK_{\mathcal{I}} +\tau_pK_{\mathcal{I}}^2)$ flops. The calculation includes the cost of computing the SVD, $\widehat{\bar{\mathbf{S}}}_{l-1}$ and $\mathbf{Q}_l$, the products ($\widehat{\bar{\mathbf{S}}}_l^{\circ}$ and $\mathbf{Q}_l^H$), and the addition of the matrices ($\widehat{\bar{\mathbf{S}}}_{l-1}$ and $\widehat{\bar{\mathbf{S}}}_l^{\circ}\mathbf{Q}_l^H$) in~\eqref{RotnAvg}.

\subsection{Method 2: Sequential Accumulation of Gramians}
This method is based on accumulation of Gramians of the residual signal and was also proposed in \cite{shaik2023distributed} (for a single OoS source). First observe that the columns of $\widehat{\bar{\mathbf{S}}}$ are the $K_{\mathcal{I}}$ dominant eigen-vectors of the Gramian of $\mathbf{Z}\mathbf{\Psi}$ i.e.,
\begin{equation}\label{eqn:Gram}
	\boldsymbol{\Psi}^H\mathbf{Z}^H\mathbf{Z}\boldsymbol{\Psi} = \sum_{l=1}^{L}(\mathbf{Z}_l\boldsymbol{\Psi})^H(\mathbf{Z}_l\boldsymbol{\Psi}).
\end{equation}
We accumulate the Gramians in \eqref{eqn:Gram} sequentially in the network by add-and-forward through APs. Specifically, each AP computes the sum of its local Gramian and the accumulated Gramian received from previous APs and forwards to the next AP. Finally, AP $L$ forwards the final Gramian to the CPU, and then the CPU computes the estimate of $\bar{\mathbf{S}}$ by taking the $K_{\mathcal{I}}$ dominant eigen-vectors of the Gramian in \eqref{eqn:Gram} and forwards the estimate to all APs. Then all APs compute the estimate of the OoS source channels, $\mathbf{G}_l$, using \eqref{eqn:intChanEst}.\footnote{The Gramian-accumulation method is equivalent to the centralized algorithm. Analyzing the Procrustes algorithm theoretically is challenging, and we have to leave that  for potential future work.}
The fronthaul load with this method is $(\tau_p - K)^2$ real symbols independent of the number of OoS sources. The cost of this approach is approximately $\mathcal{O}( (N+K_{\mathcal{I}})\tau_p^2)$ flops. The cost includes: computing the Gramian, addition of two Gramians, and  computing $K_{\mathcal{I}}$ dominant eigenvectors.\vspace{-4mm}

\section{Uplink/Downlink Payload Transmission}
In this section, we use the OoS interference estimates we described in the previous section for uplink and downlink processing in the presence of OoS sources. 

\subsection{Uplink: Suppression of OoS Interference}
In the uplink, we provide two ways of estimating the serving users in the presence of multiple OoS users. The first one is sequential least squares and the second one is distributed zero-forcing (ZF).

To begin, consider the signal that AP $l$ receives in a symbol period, given in \eqref{ulSigMod}. The APs use the channel estimates and the estimates of the OoS interference to suppress the OoS interference and also estimate the data of the UEs. During the uplink phase, the APs treat the OoS sources as extra fictitious users and eventually discard the corresponding detected "data symbols".

\subsubsection{Sequential LS}
For a sequential setup, methods that involve each AP forwarding local estimate to the consecutive AP which then computes a improved version of the estimate based on its local data are suitable as they keep fronthaul signaling in the links between APs constant. One such method is a sequential LS. We can implement sequential LS combining of the desired signal and the interfering signal at AP $l$ as
\begin{equation} \label{SeqLS}
	\begin{bmatrix}
		\widehat{\mathbf{x}}_l^{\rm ul}\\ \widehat{\mathbf{s}}_l
	\end{bmatrix} =  \begin{bmatrix}
		\widehat{\mathbf{x}}_{l-1}^{\rm ul}\\ \widehat{\mathbf{s}}_{l-1}
	\end{bmatrix} + \mathbf{T}_l\left(\mathbf{y}_l^{\rm ul} - \widehat{\mathbf{A}}_l  \begin{bmatrix}
		\widehat{\mathbf{x}}_{l-1}^{\rm ul}\\ \widehat{\mathbf{s}}_{l-1}
	\end{bmatrix}\right), l \in [L],
\end{equation}
where {\small$\widehat{\mathbf{A}}_l  = \begin{bmatrix}
	\widehat{\mathbf{H}}_l & \widehat{\mathbf{G}}_l
\end{bmatrix}$, $\mathbf{T}_l = \mathbf{C}_{l-1}	\widehat{\mathbf{A}}_l^H\left(\sigma^2\mathbf{I} + 	\widehat{\mathbf{A}}_l \mathbf{C}_{l-1}	\widehat{\mathbf{A}}_l^H\right)^{-1}$ and $\mathbf{C}_l = \left(\mathbf{I} - \mathbf{T}_l 	\widehat{\mathbf{A}}_l\right)\mathbf{C}_{l-1}$ }
\begin{comment}
\begin{align}
	\widehat{\mathbf{A}}_l & = \begin{bmatrix}
		\widehat{\mathbf{H}}_l & \widehat{\mathbf{G}}_l
	\end{bmatrix}\\
	\mathbf{T}_l &= \mathbf{C}_{l-1}	\widehat{\mathbf{A}}_l^H\left(\sigma^2\mathbf{I} + 	\widehat{\mathbf{A}}_l \mathbf{C}_{l-1}	\widehat{\mathbf{A}}_l^H\right)^{-1},\\
	\mathbf{C}_l &= \left(\mathbf{I} - \mathbf{T}_l 	\widehat{\mathbf{A}}_l\right)\mathbf{C}_{l-1},
\end{align}
\end{comment}
and initial values $\begin{bmatrix}
	\widehat{\mathbf{x}}_0^{\rm ul}\\ \widehat{\mathbf{s}}_0
\end{bmatrix} = \mathbf{0}$, $\mathbf{C}_{0} = \alpha \mathbf{I}$ with $\alpha$ being some large positive constant, this is to avoid biasing the estimator towards the initial estimate $\widehat{\mathbf{S}}_{0}$. Note that the matrices $\{\mathbf{C}_l\}$ are positive semi-definite as these are error covariance matrices~\cite{kay1993fundamentals}.

\subsubsection{Distributed Uplink ZF}
To implement ZF uplink combining, AP $l$ combines the received signal in \eqref{ulSigMod} with $\widehat{\mathbf{A}}_l^H$ and then adds it to the processed received signal from the previous AP as
\begin{equation}\label{seqZFul0}
	\begin{aligned}
		\bar{\mathbf{y}}_{l}^{\rm ul} &= \bar{\mathbf{y}}_{l-1}^{\rm ul} + \widehat{\mathbf{A}}^H_l\mathbf{y}_{l}^{\rm ul};\ \bar{\mathbf{y}}_{0}^{\rm ul} = \mathbf{0},
	\end{aligned}
\end{equation}
where $\mathbf{y}_{l}^{\rm ul}$ is the received signal and $\bar{\mathbf{y}}_{l}^{\rm ul}$  is the processed received signal in the uplink at AP $l$. This process continues until the last AP. The CPU receives $\bar{\mathbf{y}}^{\rm ul} = \bar{\mathbf{y}}_L^{\rm ul}$ and it makes the signal estimates as follows:
\begin{equation}\label{seqZFul}
	\begin{aligned}
		\begin{bmatrix}
			\widehat{\mathbf{x}}^{\rm ul}\\ \widehat{\mathbf{s}}
		\end{bmatrix} = \widehat{\boldsymbol{\Gamma}}^{-1}\bar{\mathbf{y}}^{\rm ul};\ \widehat{\boldsymbol{\Gamma}} \triangleq \widehat{\mathbf{A}}^H\widehat{\mathbf{A}},
	\end{aligned}
\end{equation}
where $\widehat{\mathbf{x}}^{\rm ul} $ is the estimate of the transmit signal from all UEs, and $\widehat{\mathbf{s}}$ is the estimate of signals from OoS sources (which are discarded) and $\widehat{\mathbf{A}} = [	\widehat{\mathbf{A}}_1^T,\ldots,	\widehat{\mathbf{A}}_L^T]^T$. We note that this Gramian accumulation can be used for both uplink and downlink processing. In this method, each AP needs the Gramian of the all the channels. To accomplish this in a sequential network, AP $l$ computes the Gramian of the effective channel estimate $\widehat{\mathbf{A}}_l$ and then adds it to the effective Gramian it receives from the previous AP as follows:
\begin{equation}\label{accuGrams}
	\widehat{\boldsymbol{\Gamma}}_l = \widehat{\boldsymbol{\Gamma}}_{l-1} +  \widehat{\mathbf{A}}_l^H\widehat{\mathbf{A}}_l,\ 	\widehat{\boldsymbol{\Gamma}}_0 = \mathbf{0},
\end{equation}
and then forwards this to the next AP.  The final AP forwards the final Gramian, $\widehat{\boldsymbol{\Gamma}}= \widehat{\boldsymbol{\Gamma}}_l$ to the CPU.

\subsubsection{Centralized ZF (Baseline)}
For reference, the centralized ZF/LS is given by
\begin{equation}\label{centLS}
	\begin{bmatrix}
		\widehat{\mathbf{x}}^{\rm ul}\\ \widehat{\mathbf{s}}
	\end{bmatrix} =  \widehat{\mathbf{A}}^{\dagger}\mathbf{y}^{\rm ul} ,\vspace{-2mm}
\end{equation}
where $\widehat{\mathbf{x}}$ and $\widehat{\mathbf{s}}$ are the estimates of $\mathbf{x}$ and $\mathbf{s}$, respectively, $\mathbf{y}^{\rm ul} = [\mathbf{y}_1^{{\rm ul}T},\ldots,\mathbf{y}_L^{{\rm ul}T}]^T$ is the augmented received signal from all APs, and $(\cdot)^\dagger$ is the pseudoinverse.

Among the above three methods, centralized ZF \eqref{centLS} will have superior performance, distributed ZF \eqref{seqZFul} have an equivalent performance as centralized ZF. Sequential LS in \eqref{SeqLS} may have suboptimal performance compared to other two methods. However, sequential LS does not need to compute Gramians of the channel estimates given in \eqref{seqZFul}.

\subsection{Downlink: Nullify the Signal in the OoS Direction}
In applications where we intend to protect the OoS sources or nullify the signal in the direction to them, we can use the estimate of the OoS sources channels to transmit the serving UEs signal in the nullspace of the OoS channel.

The process of nullifying the signal in a particular direction can be accomplished with the ZF precoder. But, it is challenging to implement ZF precoding exactly in a decentralized network \cite{sarajlic2019fully}. We will now provide one way to implement centralized ZF on a stripe topology. To do this, we start by considering ZF precoding of a centralized cell-free network
\begin{equation}
	\overline{\mathbf{W}} = 	\widehat{\mathbf{A}}\left(\widehat{\mathbf{A}}^H\widehat{\mathbf{A}}\right)^{-1},
\end{equation}
where $\overline{\mathbf{W}} = [\overline{\mathbf{W}}_1^T,\ldots,\overline{\mathbf{W}}_L^T]^T$ is the centralized precoder with $\overline{\mathbf{W}}_l$ being the precoder at AP $l$. Then, we can write the local precoding at AP $l$ as $\overline{\mathbf{W}}_l = 	\widehat{\mathbf{A}}_l\widehat{\boldsymbol{\Gamma}}^{-1}$, where the Gramian of the channel estimates $\widehat{\boldsymbol{\Gamma}}$ obtained at the CPU in \eqref{accuGrams} can be used. Note that the first $K$ columns of $\overline{\mathbf{W}}_l$ forms the effective transmit precoder ${\mathbf{W}}_l$ defined in \eqref{dlSigMod}.

Let $\mathbf{x}^{\rm dl} \in \complexm{K}{1}$ be the signal vector that we intend to transmit in downlink, whose $k$th entry is the signal component for UE $k\in [K]$. Then, the downlink signal model is given by
{\small
\begin{equation}\label{dlSig}
	\begin{aligned}
		\mathbf{y}^{\rm dl} &= \sum_{l=1}^{L}\mathbf{H}_l^H\overline{\mathbf{W}}_l\begin{bmatrix}
			\mathbf{x}^{\rm dl}\\ \mathbf{0}_{K_{\mathcal{I}}}
		\end{bmatrix} + \mathbf{n}_l^{\rm dl}\\
		&= \sum_{l=1}^{L}\mathbf{H}_l^H\widehat{\mathbf{A}}_l\underbrace{\widehat{\boldsymbol{\Gamma}}^{-1}\begin{bmatrix}
				\mathbf{x}^{\rm dl}\\ \mathbf{0}_{K_{\mathcal{I}}}
		\end{bmatrix}}_{\mathbf{q}}+ \mathbf{n}^{\rm dl}.
	\end{aligned}
\end{equation}}
Note that we do not send any signal in the direction of OoS sources. An important observation from \eqref{dlSig} is that the part of precoded signal component, $\mathbf{q}$, is independent of the AP index. Further, the Gramian can be accumulated at the CPU sequentially by adding and forwarding the local Gramians as in \eqref{accuGrams}. Then, in the downlink, the CPU forms the partially precoded signal, $\mathbf{q}$ and forwards to all APs. In this way, we accomplish centralized ZF on a stripe topology and effectively achieve nulling of the signal in the direction of the OoS sources.

\begin{figure}[!t]
	\centering
	%\immediate\write18{pdfcrop ./Figures/processFlowPDF.pdf}
	%\includegraphics[trim={0 1.5cm 0 1.2cm},clip, height=4cm, width=8.5cm]{processFlowPDF-crop}
	\includegraphics[trim={0 1.5cm 0 1.2cm},clip,width=1\linewidth,keepaspectratio]{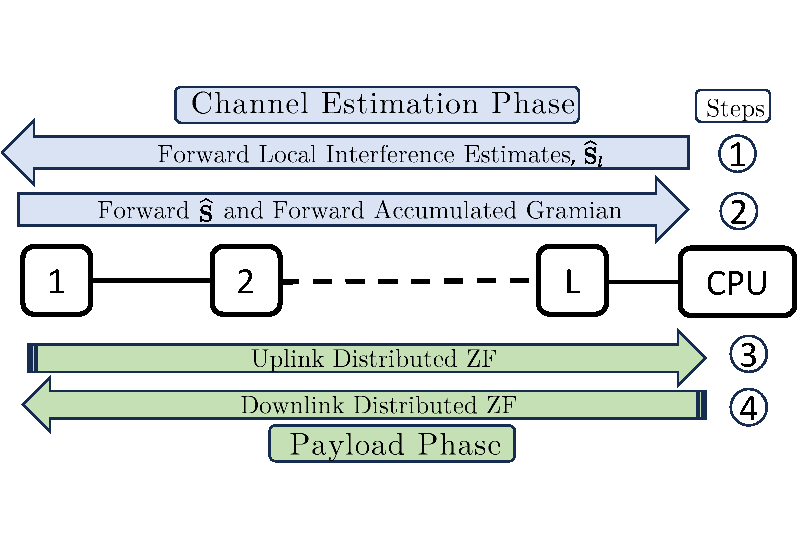}
	\caption{Overall flow of processing in pilot and data phase}
	\label{fig:procflow}\vspace{-3mm}
\end{figure}

A flowchart of the overall processing is shown in Fig.~\ref{fig:procflow}. The first step involves estimating the OoS interfering signal (note that the order of the processing is reversed, i.e., from $L$ to $1$ to reduce latency). The second step involves computing local OoS channel estimates and forwarding accumulated Gramians. In the third step, we perform \eqref{seqZFul0} and in the final step we compute $\mathbf{q}$ given in \eqref{dlSig}. \vspace{-3mm}

% [trim={left bottom right top},clip]
\section{Numerical Results}
In this section, we evaluate the performance of the proposed methods through numerical simulations. We consider a $500 \times 500$ square meter area with APs equally spaced on the border. We deploy UEs uniformly within the concentric square area with $10 {\rm m}$ gap from the border. We consider the standard Rayleigh fading channel model i.e., $\mathbf{h}_{kl}\sim \mathcal{CN}(\mathbf{0},\beta_{kl} \mathbf{I})$ where $\beta_{kl}$ is the large-scale fading coefficient and we take it to be $\beta_{kl}~[{\rm dB}] = -30.5 - 36.7\log_{10}(d_{kl}/1 {\rm m})$, where $d_{kl}$ is the distance between AP $l$ and UE $k$. We consider that APs are at vertical height of $5~{\rm m}$. At each AP, we consider a uniform linear array with half-wavelength antenna spacing. We consider the number of APs to be $L=4$, the number of antennas per AP $N=4$, and the number of UEs $K=5$. We consider $\tau_p = 50$ and $\tau_c = 200$ channel uses.

For the performance analysis we consider the uplink payload phase and we use QPSK modulated data symbols. We employ a nearest-point detector on the estimated signals i.e., for \eqref{centLS} for centralized processing and \eqref{seqZFul} or \eqref{SeqLS} for the sequential network. The OoS sources signals are drawn from a Gaussian distribution and assume have transmit SNR of $-3~{\rm dB}$. Methods that are compared are:
\begin{itemize}
	\item {\bf No interference suppression}: Use \eqref{centLS} to estimate data of serving UEs without suppressing OoS interference.
	\item {\bf Local processing}: Estimate UEs payload signal using only local estimates of the OoS interference. The cost of this method is approximately $\mathcal{O}(NK_{\mathcal{I}}\tau_p)$ flops.
	\item {\bf Sequential Procrustes method}: This is the proposed distributed OoS suppression method using the orthogonal Procrustes algorithm in \eqref{procrustesProb1}.
	\item {\bf Sequential Gramians based method}: This is the accumulation of Gramians method to estimate OoS interference as in \eqref{eqn:Gram}, which has identical performance to centralized processing.
	\item {\bf Centralized genie detector}: This is the baseline method with perfect knowledge of all channels (both serving UEs and OoS sources) i.e., $\{\mathbf{H}_l,\mathbf{G}_l; l \in [L]\}$.
\end{itemize}

\begin{figure}[!t]
	\centering
	\includegraphics[width=1\linewidth,keepaspectratio]{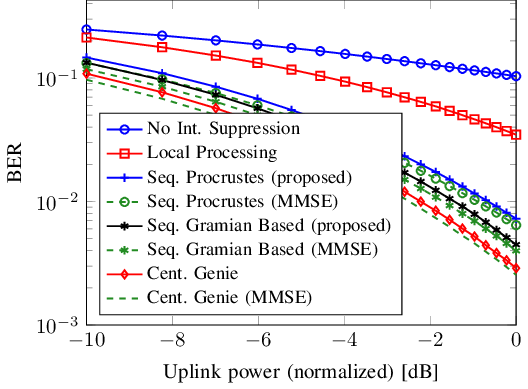}
	\caption{Performance of the proposed algorithm with $K_{\mathcal{I}} = 2$}
	\label{plot1}\vspace{-5mm}
\end{figure}

Fig. \ref{plot1} shows the results. There is a notable degradation in performance if the network do not suppress the interference. The Gramian-based method has superior performance among the distributed algorithms as it has equivalent performance to that of centralized processing (also to that of distributed ZF uplink combining). Local processing have the worst performance although it has the lowest fronthaul load requirement. From the simulation, we note that the proposed sequential Procrustes method offers very good trade-off in terms of performance and fronthaul load requirements. However, when $K_{\mathcal{I}}>N$, the performance of the Procrustes method degrades and in this situation the Gramian-based method should be preferred. The curves in green represent the scenarios where we employ minimum-mean-square-error (MMSE) channel estimation and uplink combining, 
considering the Rayleigh fading priors on the channels while not relying on any statistical information on the   OoS interferer channels or signals.
The relative performance of the different methods remains substantially the same when MMSE is used. Besides the UEs performance, the proposed sequential Procrustes algorithm exhibits a complexity that is in between that of centralized processing and local processing, with the latter being the least complex. 
In the downlink, two possible metrics of performance could be:  the QoS of the serving users, and the  received signal energy (which should be minimal) at the OoS sources. Due to  space limitations, we   include the performance evaluation only for the uplink scenario.
	
\section{Conclusion}
This paper proposed decentralized algorithms to estimate the OoS interference. We presented two applications where the estimate of OoS interference is used to cancel the interference in the uplink and to suppress the transmit signal in the direction of OoS in the downlink. The proposed distributed Procrustes method leverages the fact that OoS interference is the same across all the APs. We demonstrated through simulation results that the proposed algorithm has performance close to that of centralized implementation but less fronthaul load. Specifically, the proposed Procrustes based method has a fronthaul load that is linear in the number of UEs, in contrast to the quadratic proportionality with a centralized approach. As a possible topic for future work, one could explore  
 suppression of OoS interference when over-the-air computation is used for the communication among APs or between the APs and the CPU  \cite{dai2023learning,shaik2023resource}. \vspace{-4mm}
\bibliographystyle{IEEEtran}
\bibliography{WCL2024-0565.R1.bib}

\end{document}